# A Survey on Cloud Security Issues and Techniques

Garima Gupta[1], P.R.Laxmi[2] and Shubhanjali Sharma[3]

[1]Department of Computer Engineering, Government Engineering College, Ajmer
Guptagarima09@gmail.com
[2]Department of Computer Engineering, Government Engineering College, Ajmer
laxmigweca@gmail.com
[3]Department of Computer Engineering, Government Engineering College, Ajmer
shubhanjali.sharma26@gmail.com

## Abstract:

*Today, cloud computing is an emerging way of computing in computer science. Cloud computing is a set of resources and services that are offered by the network or internet. Cloud computing extends various computing techniques like grid computing, distributed computing. Today cloud computing is used in both industrial field and academic field. Cloud facilitates its users by providing virtual resources via internet. As the field of cloud computing is spreading the new techniques are developing. This increase in cloud computing environment also increases security challenges for cloud developers. Users of cloud save their data in the cloud hence the lack of security in cloud can lose the user's trust.*

*In this paper we will discuss some of the cloud security issues in various aspects like multi-tenancy, elasticity, availability etc. the paper also discuss existing security techniques and approaches for a secure cloud. This paper will enable researchers and professionals to know about different security threats and models and tools proposed.*

## 1. Introduction

Cloud computing is another name for Internet computing. The definition of cloud computing provided by National Institute of Standards and Technology (NIST) says that: "Cloud computing is a model for enabling on-demand and convenient network access to a shared pool of configurable computing resources (e.g., networks, servers, storage applications and services) that can be rapidly provisioned and released with minimal management effort or service provider interaction[9]. For some it is a paradigm that provides computing resources and storage while for others it is just a way to access software and data from the cloud. Cloud computing is popular in organization and academic today because it provides its users scalability, flexibility and availability of data. Also cloud computing reduces the cost by enabling the sharing of data to the organization. Organization can port their data on the cloud so that their shareholders can use their data. Google apps is an example of cloud computing.

However Cloud provides various facility and benefits but still it has some issues regarding safe access and storage of data. Several issues are there related to cloud security as: vendor lock-in, multi-tenancy, loss of control, service disruption, data loss etc. are some of the research problems in cloud computing [2]. In this paper we analyze the security issues related to cloud computing model. The main goal is to study different types of attacks and techniques to secure the cloud model.

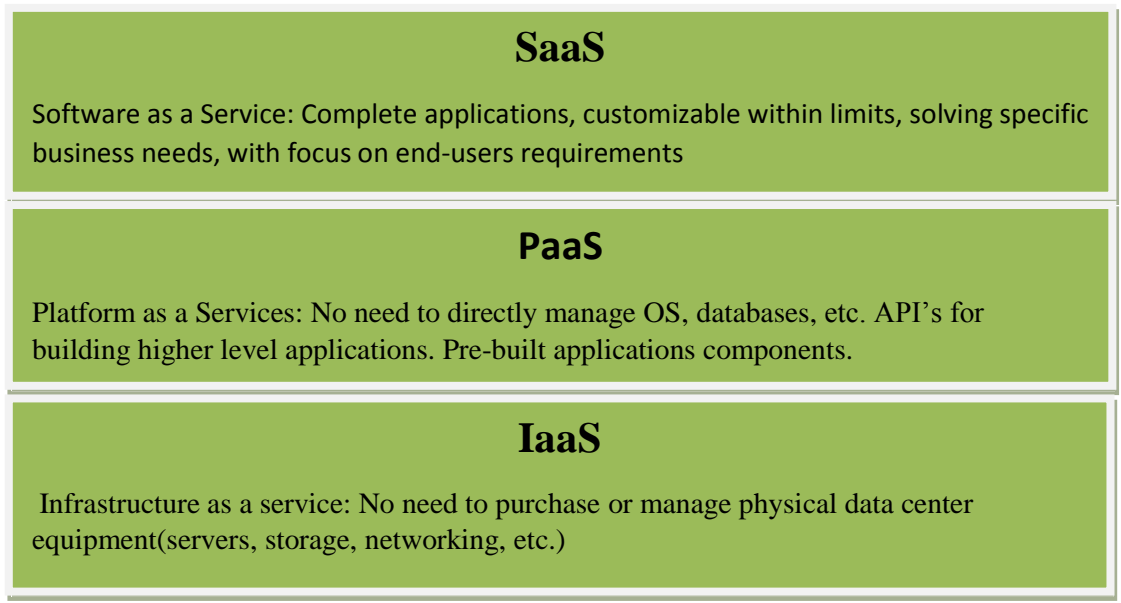

**Figure 1 .** Layers of Cloud Computing

## 2. Cloud security issues

Organization uses various cloud services as IaaS, PaaS, SaaS and the models like public, private, hybrid. These models and services has various cloud security issues. Each service model is associated with some issues. Security issues are considered in two views first in the view of service provider who insures that services provided by them should be secure and also manages the customer's identity management. Other view is customer view that ensures that service that they are using is secure enough.

### 2.1 Multi-tenancy

A cloud model is built for reasons like sharing of resources, memory, storage and shared computing [2]. Multi-tenancy provides efficient utilization of resources, keeping cost lower. It implies sharing of computational resources, services storage and application with other tenants residing on same physical/logical platform at provider's premises. Thus it violates the confidentiality of data and results in leakage of information and encryption and increase the possibility of attacks.

### 2.2 Elasticity

Elasticity is defined as the degree to which a system is able to adapt to workload changes by provisioning and deranged resources in an autonomic manner, such that the available resources match the current demand at any time as closely as possible. Elasticity implies scalability. It says that consumers are able to scale up and down as needed. This scaling enables tenants to use a resource that is assigned previously to other tenant. However this may lead to confidentiality issues.

### 2.3  Insider attacks

Cloud model is a multitenant based model that is under the provider's single management domain. This is a threat that arises within the organization. There are no hiring standards and providers for cloud employees [1]. So a third party vendor can easily hack the data of one organization and may corrupt or sell that data to other organization.

### 2.4  Outsider attacks

This is the one of the major concerning issue in an organization because it releases the confidential information of an organization in open. Clouds are not like a private network, they have more interfaces than private network. So hackers and attackers have advantage of exploiting the API, weakness and may do a connection breaking [1] .These attacks are less harmful than the insider attacks because in the later we sometimes unable to identify the attack.

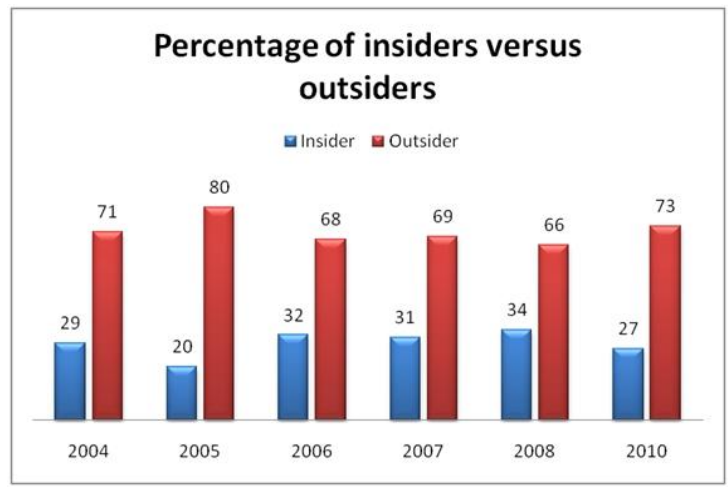

**Figure 2 .** Percentage of Insiders versus Outsiders [1]

### 2.5  Loss of control

Cloud uses a location transparency model by which it enable organizations to unaware about the location of their services and data. Hence provider can host their services from anywhere in the cloud. In this case organization may lose their data and possibly they are not aware about security mechanism put in place of the provider.

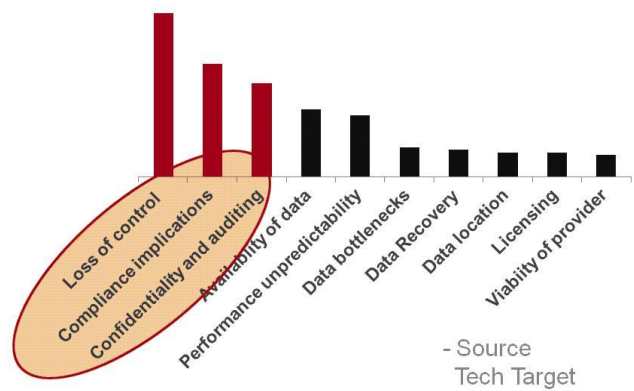

**Figure 3 .** Loss of Control over Data [1]

### 2.6 Data Loss

As in cloud, there are multiple tenants, data integrity and safety could not be provided. Data loss can results in financial, customer count loss for an organization. An important example of this can be updating and deletion of data without having any backup of that data.

### 2.7 Network security

#### 2.7.1 Man in middle attack:-

In this attack, attacker makes an independent connection and communicates with the cloud user on its private network where all control is in the hand of attacker.

#### 2.7.2 Distributed denial of service attacks: -

 In DDOS attack, servers and networks are brought down by a huge amount of network traffic and users are denied the access to a certain Internet based Service.[3]

#### 2.7.3 Port scanning:-

 Port is a place from where information exchange takes place. Port scanning is taking place when subscriber configures the group. Port scanning is done automatically when you configure the internet so this violates the security concerns [3].

### 2.8 Malware Injection Attack Problem

In cloud computing, a lot of data is transferred between cloud provider and consumer, there is a need of user authentication and authorization [10]. When the data is transferred between cloud provider and user, attacker can introduce malicious code into it. As a result, the original user may have to wait until the completion of the job that was maliciously introduced.

### 2.9 Flooding Attack Problem

In cloud, there is a no. of servers that communicate with one another and transfer data. The requests is processed, the requested jobs are authenticated first, but this authentication requires a lot of CPU utilization, memory and finally due to these server is overloaded and it passes its offload to other server[10]. By all this the usual processing of system is interrupted, and the system is flooded.

## 3. Techniques to secure data in cloud

### 3.1 Authentication and Identity

Authentication of users and even of communicating systems is performed by various methods, but the most common is cryptography [8]. Authentication of users takes place in various ways like in the form of passwords that is known individually, in the form of a security token, or in the form a measurable quantity like fingerprint. One problem with using traditional identity approaches in a cloud environment is faced when the enterprise uses multiple cloud service providers (CSPs)[8]. In such a use case, synchronizing identity information with the enterprise is not scalable. Other problems arise with traditional identity approaches when migrating infrastructure toward a cloud-based solution.

### 3.2 Data Encryption

If you are planning to store sensitive information on a large data store then you need to use data encryption techniques. Having passwords and firewalls is good, but people can bypass them to access your data. When data is encrypted it is in a form that cannot be read without an encryption key. The data is totally useless to the intruder. It is a technique of translation of data into secret code. If you want to read the encrypted data, you should have the secret key or password that is also called encryption key.

### 3.3 Information integrity and Privacy

Cloud computing provides information and resources to valid users. Resources can be accessed through web browsers and can also be accessed by malicious attackers [2]. A convenient solution to the problem of information integrity is to provide mutual trust between provider and user. Another solution can be providing proper authentication, authorization and accounting controls so the process of accessing information should go through various multi levels of checking to ensure authorized use of resources [2]. Some secured access mechanisms should be provided like RSA certificates, SSH based tunnels.

### 3.4 Availability of Information(SLA)

Non availability of information or data is a major issue regarding cloud computing services. Service Level agreement is used to provide the information about whether the network resources are available for users or not. It is a trust bond between consumer and provider [2].An way to provide availability of resources is to have a backup plan for local resources as well as for most crucial information. This enables the user to have the information about the resources even after their unavailability.

### 3.5 Secure Information Management

It is a technique of information security for a collection of data into central repository. It is comprised of agents running on systems that are to be monitored and then sends information to a server that is called "Security Console". The security console is managed by admin who is a human being who reviews the information and takes actions in response to any alerts. As the cloud user base, dependency stack increase, the cloud security mechanisms to solve security issues also increase, this makes cloud security management much more complicated. It is also referred as a Log Management. Cloud providers also provide some security standards like PCI DSS, SAS 70[2]. Information Security Management Maturity is another model of Information Security Management System.

### 3.6 Malware-injection attack solution

This solution creates a no. of client virtual machines and stores all of them in a central storage. It utilizes FAT (File Allocation Table) consisting of virtual operating systems[10]. The application that is run by a client can be found in FAT table. All the instances are managed and scheduled by Hypervisor. IDT (Interrupt Descriptor Table) is used for integrity checking.

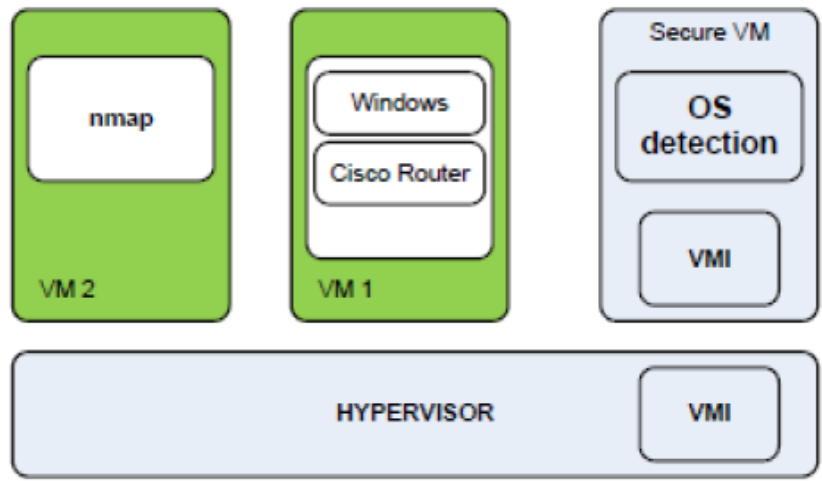

**Figure 4 .** Malware-Injection attack solution [10]

### 3.7 Flooding Attack Solution

All the servers in cloud are considered as a fleet of servers. One fleet of server is considered for system type requests, one for memory management and last one for core computation related jobs. All the servers in fleet can communicate with one another. When one of the server is overloaded, a new server is brought and used in the place of that server and an another server that is called name server has all the record of current states of servers and will be used to update destinations and states. Hypervisor can be used for managing jobs[10]. Hypervisor also do the authorization and authentication of jobs. An authorized customer's request can be identified by PID. RSA can also be used to encrypt the PID.

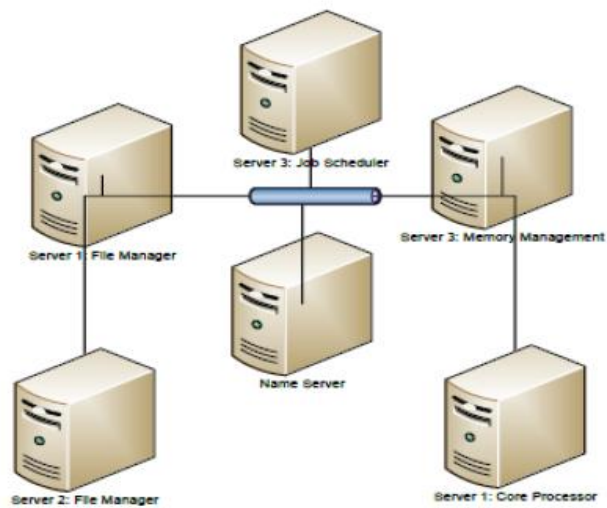

**Figure 5 .** Flooding Attack solution [10]

## 4. Cloud computing Security Standards

Standards for security define procedure and processes for implementing a security program. To maintain a secure environment, that provides privacy and security some specific steps are performed by applying cloud related activities by these standards. A concept called "Defence in Depth" is used in cloud to provide security [3]. This concept has layers of defence. In this way, if one of the systems fails, overlapping technique can be used to provide security as it has no single point of failure. Traditionally, endpoints have the technique to maintain security, where access is controlled by user.

### 4.1 Security Assertion Markup Language (SAML)

SAML is basically used in business deals for secure communication between online partners. It is an XML based standard used for authentication, authorization among the partners. SAML defines three roles: the principal (a user), a service provider (SP) and an identity provider (IDP) [3]. SAML provides queries and responses to specify user attributes authorization and authentication information in XML format. The requesting party is an online site that receives security information.

### 4.2 Open Authentication (OAuth)

It is a method used for interacting with protected data. It is basically used to provide data access to developers. Users can grant access to information to developers and consumers without sharing of their identity [3]. OAuth does not provide any security by itself in fact it depends on other protocols like SSL to provide security.

### 4.3 OpenID

OpenID is a single-sign-on (SSO) method. It is a common login process that allows user to login once and then use all the participating systems [3]. It does not based on central authorization for authentication of users.

### 4.4 SSL/TLS

TLS is used to provide secure communication over TCP/IP. TLS works in basically three phases: In first phase, negotiation is done between clients to identify which ciphers are used. In second phase, key exchange algorithm is used for authentication [3]. These key exchange algorithms are public key algorithm. The final and third phase involves message encryption and cipher encryption.

## 5. Conclusion

This paper describes some of the cloud concepts and demonstrates the cloud properties such as scalability, platform independent, low-cost, elasticity and reliability. Although there are various security challenges in cloud computing but in this paper, we have discussed some of them and also the techniques to prevent them, they can be used to maintain the secure communication and remove the security problems. This survey is basically done to study all the problems like attacks, data loss and unauthenticated access to data and also the methods to remove those problems. As the cloud computing is dynamic and complex, the traditional security solutions provided by cloud environment do not map well to its virtualized environments. Organization such as Cloud Security Alliance (CSA) and NIST are working on cloud computing security. In this paper we have discussed a few security approaches but several other approaches are also there that are in the process. Some standards are also specified which can be used to maintain secure communication and security in a cloud as many systems communicate in it and perform operations.

**Authors**

Garima Gupta Completed B.Tech in CSE from GVSET Jaipur in 2011 under Rajasthan Technical University. She is currently pursuing M.Tech from Govt Engineering College Ajmer. Her research area includes Networking and Cloud Computing.

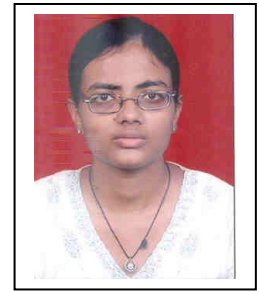

P.R. Laxmi Completed B.Tech in CSE from GWECA Ajmer in 2011 under Rajasthan Technical University. She is currently pursuing M.Tech from Govt Engineering College Ajmer. Her research area includes Networking and Cloud Computing.

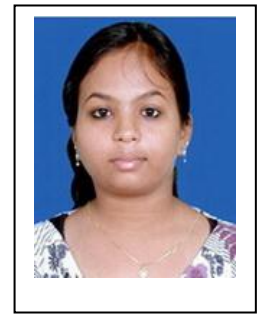

Shubhanjali Sharma Completed B.Tech in IT from GWECA Ajmer in 2011 under Rajasthan Technical University. She is currently pursuing M.Tech from Govt Engineering College Ajmer. Her research area includes Networking and Cloud Computing and Security.

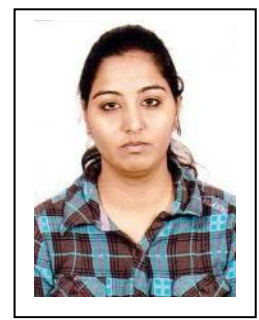